%% file: paper616.tex
\documentclass[runningheads]{llncs}
\usepackage{amsmath}
\usepackage{amsfonts}
\usepackage{xcolor}
\usepackage{graphicx}
\usepackage{hyperref}
\usepackage{fontawesome}
\usepackage{ifthen}
\usepackage{booktabs}
\usepackage{multirow}
\usepackage{subcaption}
\usepackage[referable]{threeparttablex}
\usepackage[ruled,vlined]{algorithm2e}
\usepackage{verbatim}
\usepackage{xcolor}
\usepackage{tikz}
\SetKwComment{Comment}{$\triangleright$\ }{}
\usepackage{dcolumn} 
\newcolumntype{d}[1]{D{.}{.}{#1}}
\newboolean{doc_anonymous}   
\newboolean{doc_appendix}

\DeclareMathOperator*{\argmin}{arg\,min}

\begin{document}
\title{Divide-and-Rule: Self-Supervised Learning for Survival Analysis in Colorectal Cancer}
\titlerunning{Divide-and-Rule: Self-Supervised Learning for Survival Analysis in CRC}

\setboolean{doc_anonymous}{false}   
\setboolean{doc_appendix}{true} 

\ifthenelse{\boolean{doc_anonymous}}
{
\author{Anonymous \textsuperscript{\faEnvelopeO}}
\authorrunning{Anonymous}
\institute{Anonymous Organization\\
\email{***@***.***}\\
\url{www.***.***}}
}
{
\author{Christian Abbet\inst{1} \textsuperscript{\faEnvelopeO}
\and Inti Zlobec\inst{4} \and \\
 Behzad Bozorgtabar\inst{1, 2,3} \and Jean-Philippe Thiran\inst{1, 2,3}}
\authorrunning{C. Abbet et al.}
\institute{Signal Processing Laboratory 5, EPFL, Lausanne, Switzerland \email{\{firstname.lastname\}@epfl.ch}
\and {Department of Radiology, Lausanne University Hospital, Lausanne, Switzerland}
 \and {Center of Biomedical Imaging, Lausanne, Switzerland}
\and TRU – Translational Research Unit, Bern, Switzerland
}
}

\maketitle              

\input{files/abstract.tex}

\input{files/introduction.tex}

\input{files/method.tex}

\input{files/results.tex}

\input{files/conclusion.tex}


\bibliographystyle{splncs04}
\bibliography{paper616.bib}

\appendix
\input{files/appendix.tex}

\end{document}

%% file: files/abstract.tex
\begin{abstract}
With the long-term rapid increase in incidences of colorectal cancer (CRC), there is an urgent clinical need to improve risk stratification. The conventional pathology report is usually limited to only a few histopathological features. However, most of the tumor microenvironments used to describe patterns of aggressive tumor behavior are ignored. In this work, we aim to learn histopathological patterns within cancerous tissue regions that can be used to improve prognostic stratification for colorectal cancer. To do so, we propose a self-supervised learning method that jointly learns a representation of tissue regions as well as a metric of the clustering to obtain their underlying patterns. These histopathological patterns are then used to represent the interaction between complex tissues and predict clinical outcomes directly. We furthermore show that the proposed approach can benefit from linear predictors to avoid overfitting in patient outcomes predictions. To this end, we introduce a new well-characterized clinicopathological dataset, including a retrospective collective of 374 patients, with their survival time and treatment information. Histomorphological clusters obtained by our method are evaluated by training survival models. The experimental results demonstrate statistically significant patient stratification, and our approach outperformed the state-of-the-art deep clustering methods. 

\keywords{Self-supervised learning \and Histology \and Survival analysis \and Colorectal cancer.}
\end{abstract}

%% file: files/introduction.tex
\section{Introduction}
Colorectal cancer is the third leading cause of cancer-related mortality worldwide. Five-year survival rates are low, at 60$\%$. Although standard histopathological of cancer reporting based on features such as staging and grading identifies patients with a potentially worse outcome to therapy, there is still an urgent need to improve risk stratification. Pathologists typically limit their reporting of colorectal cancers to approximately ten features, which they describe as single elements in their report (e.g., depth of invasion, pT; lymph node metastasis, etc.). However, the histopathological (H\&E) slide is a “snapshot” of all occurring tumor-related processes, and their interactions may hold a wealth of information that can be extracted to help refine prognostication. These slides can then be digitized and used as input for computational algorithms to help support pathologists in their decision-making. The distribution of tissue types within the slide, the proximity of cell types or tissue components, and their spatial arrangement throughout the tissue can identify new patterns not previously detectable to the human eye alone.

Few studies have performed unsupervised clustering of whole slide images (WSIs) based on patch descriptors. They have been used to address the problem of image segmentation \cite{moriya2018unsupervised} or latent space clustering \cite{dercksen2019dealing,fouad2017unsupervised}. Among DL-based survival models, a recent study \cite{kather2019predicting} used a supervised CNN for end-to-end classification of tissues to predict the survival of patients with colorectal cancer. 
Similar to our approach, several recent works have proposed unsupervised methods \cite{muhammad2019unsupervised,zhu2017wsisa,li2018graph} for slide-level survival analysis. In \cite{zhu2017wsisa}, one of the first unsupervised approaches, DeepConvSurv has been proposed for survival prediction based on WSIs. More recently, DeepGraphSurv \cite{li2018graph} has been presented to learn global topological representations of WSI via graphs. However, they heavily relied on noisy compressed features from a pre-trained VGG network. Recently, self-supervised representation learning methods \cite{he2020momentum,zhuang2019local,chen2020simple} have been proposed to utilize the pretext task for extracting generalizable features from the unlabeled data itself. Therefore, the dataset does not need to be manually labeled by qualified experts to solve the pretext task. 

\subsubsection{Contributions.} In this work, we propose a new approach to learn histopathological patterns through self-supervised learning within each WSI. Besides, we present a novel way to model the interaction between tumor-related image regions for survival analysis and tackle the inherent overfitting problem on tiny patient sets. To this end, we take advantage of a well-characterized, retrospective collective of 374 patients with clinicopathological data, including survival time and treatment information. H\&E slides were reviewed, and at least one tumor slide per patient was digitized. To accelerate research we have made our code and trained models publicly available on GitHub.\footnote{https://github.com/christianabbet/DnR}

%% file: files/method.tex
\section{Method}
\label{sec:method}
We first introduce our self-supervised image representation (Sec. \ref{subsec:SelfSupervised}) for the cancerous tissue area identified by our region of interest (RoI) detection scheme (Sec. \ref{subsec:RoIDetection}). Then, we propose our deep clustering scheme and baseline algorithms in Sec. \ref{subsec:ProposedApproach} and Sec. \ref{subsec:Algorithm Baselines}, respectively. The clustering approach's usefulness is assessed by conducting survival analysis (Sec. \ref{subsec:SurvivalAnalysis}) to measure if the learned clusters can contribute to disease prognostication. Finally, we discuss our implementation setup and experimental results in Sec. \ref{sec:experimentalResults}.

\subsection{RoI Detection}
\label{subsec:RoIDetection}
\begin{figure}[t]
    \centering
    \includegraphics[width=\textwidth]{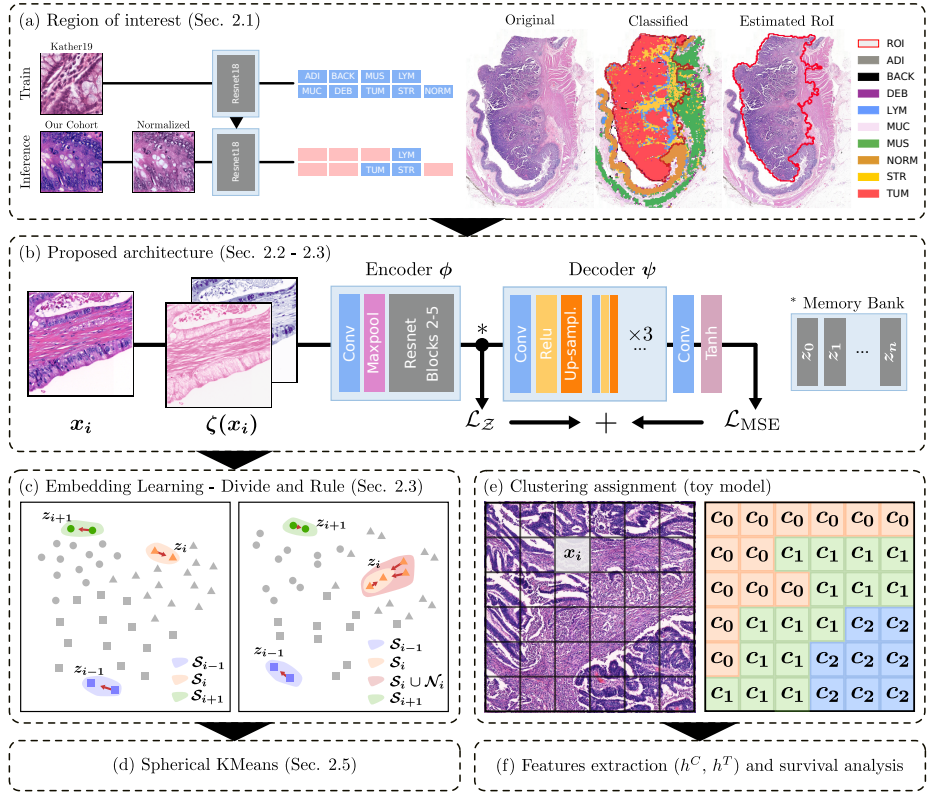}
    \caption{\textbf{The pipeline of the proposed approach.} Estimation of the region of interest (a),  learning of the embedding space (b-c), fitting of the cluster, assignment of all patient patches, and survival analysis (d-f).}
    \label{fig:model_pipeline}
\end{figure}

Our objective is to learn discriminative patterns of unhealthy tissues of patients. However, WSI does not include information about the cancerous regions or the location of the tumor itself. Therefore, we seek a transfer learning approach for the classification of histologic components of WSIs. To do so, we choose to use the dataset presented in \cite{kather19dataset} to train a classifier to discriminate relevant areas. The dataset is composed of 100K examples of tissue from CRC separated into nine different classes. For our task, we choose to retain three classes: lymphocytes (LYM), cancer-associated stroma (STR), and colorectal adenocarcinoma epithelium (TUM) that show the discriminative evidence for the class-of-interest and have been approved by the pathologist. Note that the presence of a large number of lymphocytes around the tumor is an indication of the immune reaction and, therefore, possibly linked to a higher survival score. We first train our classifier with the ResNet-18 backbone \cite{he16}. Then we use the stain normalization approach proposed in \cite{macenko09stain} to match the color space of the target domain and prevent the degradation of the classifier on transferred images. An example of RoI estimation is presented in Fig. \ref{fig:model_pipeline}. Such a technique allows us to discard a large part of the healthy tissue regions.

\subsection{Self-Supervised Representation Learning}
\label{subsec:SelfSupervised}
In this paper, we propose a self-supervised transfer colorization scheme to learn a more meaningful feature representation of the tissues and reduce the requirement for intensive tissue labeling. Unsupervised learning methods such as autoencoder trained by minimizing reconstruction error tend to ignore the underlying structure of the image as the model usually learns the distribution of the color space. To avoid this issue, we use colorization learning as a proxy task. As the input image, we convert the original unlabeled image through mapping function $\zeta(x)$ to a two-channel image (hematoxylin and eosin) that describes the nuclei and amount of extracellular material, respectively. To sidestep the memory bottleneck, we represent the WSI as a set of adjacent/overlapping tiles (image patches) $\left \{x_i \in \mathcal{X} \right \}^N_{i=1}$. 

We define a function $\zeta : \mathcal{X} \rightarrow \mathcal{X}^{HE}$ that converts the input images to their HE equivalent \cite{macenko09stain,vahadane16stain}. Then, we train a convolutional autoencoder (CAE) to measure the per-pixel difference between transformed image(s) and input image(s) using MSE loss:

\begin{equation}
    \min\limits_{\phi, \psi} \mathcal{L}_{\text{MSE}}= \min\limits_{\phi, \psi} \left \| x- \psi \circ \phi \circ \zeta \left ( x \right )\right \|_{2}^{2}
    \label{eq:autoencoder_selfsup}.
\end{equation}


The encoder $\phi: \mathcal{X}^{HE} \rightarrow \mathcal{Z}$ is a convolutional neural network that maps an input image to its latent representation $\mathcal{Z}$. The decoder $\psi: \mathcal{Z} \rightarrow \mathcal{X}$ is an up-sampling convolutional neural network that reconstructs the input image given a latent space representation. As a result, we use a single input branch to take into account the tissue's structural aspect.


\subsection{Proposed Divide-and-Rule Approach}
\label{subsec:ProposedApproach}
The principle behind our self-supervised learning approach is to represent image patches based on their spatial proximity in the feature space, meaning any two adjacent image patches (positive pairs) are more likely to be close to each other in the feature space $\mathcal{Z}$ than two distant patches (negative pairs). Such characteristics are met for overlapping patches as they share similar histomorphological patterns. We let $\mathcal{S}_i$ denote the set of patches that overlap with patch $i$ spatially. Besides, we can assume that image patches in which their relative distances are smaller than a proximity threshold in the feature space should share common patterns. We define $\mathcal{N}_i$ as the set of top-$k$ patches that achieve the lowest cosine distance to the embedding $z_i$ of the image patch $i$. 

Firstly, we initialize the network parameters using the self-supervised reconstruction loss in Eq. \ref{eq:autoencoder_selfsup}. Then, for each patch embedding $i$, we label its overlapping set of patches $\mathcal{S}_i$ as similar patches (positive pairs). Otherwise, we consider any distant patches as a negative pair, whose embeddings should be scattered. Motivated by \cite{wu2018unsupervised}, we use a variant of the cross-entropy to compute the instance loss (Eq.~\ref{eq:divide}):

\begin{equation}
    \mathcal{L}_{\text{Divide}}= -
    \sum_{i \in \mathcal{B}_{\text{inst}}} \log{(\sum_{j \in S_i} p\left( j \mid i \right))}\mathrm{,}\quad  p\left( j \mid i \right) = \frac{\exp{(z_{j}^{\top} z_{i}/\tau )}}{\sum_{k=1}^N \exp{(z_{k}^{\top} z_{i}/\tau )}}.
    \label{eq:divide}
\end{equation}

where $\tau \in \left ] 0,1 \right ]$ is the temperature parameter and $\mathcal{B}_{\text{inst}}$ denotes the set of samples in the mini-batch.  

Secondly, we jointly optimize the training of network with reconstruction loss and a Rule loss $\mathcal{L}_{\text{Rule}}$ that takes into account the similarity of different images in the feature space (Eq. \ref{eq:rule}). We gradually expand the vicinity of each sample to select its neighbor samples. If samples have high relative entropy, they are dissimilar and should be considered as individual classes, $z \in \mathcal{B}_{\text{inst}}$. On the contrary, if samples have low relative entropy with their neighbors, they should be tied together, $z \in \mathcal{Z} \backslash \mathcal{B}_{\text{inst}}$. In practice, the entropy acts as a threshold to decide a boundary between close and distant samples and is gradually increased during training such that we go from easy samples (low entropy) to hard ones (high entropy). Finally, the proposed training loss, $\mathcal{L}_{\text{DnR}}$, joins the above losses with a weighting term $\lambda$ (see Eq. \ref{eq:an_dnr}):

\begin{equation}
    \mathcal{L}_{\text{Rule}} = - \sum_{i \in \mathcal{Z}\backslash\mathcal{B}_{\text{inst}}} \log{( \sum_{j \in \mathcal{S}_i \cup \mathcal{N}_i} p\left( j \mid i \right))}.
    \label{eq:rule}
\end{equation}

\begin{equation}
    \min\limits_{\phi, \psi} \mathcal{L}_{\text{DnR}} = \min\limits_{\phi, \psi} \mathcal{L}_{\text{MSE}} + \lambda \min\limits_{\phi} [ \mathcal{L}_{\text{Divide}} + \mathcal{L}_{\text{Rule}}].
    \label{eq:an_dnr}
\end{equation}

\subsubsection{Dictionary Learning.} 
Measuring similarities between samples requires the computation of features in the entire dataset for each iteration. The complexity grows as a function of the number of samples in the dataset. To avoid this, we use a memory bank, where we keep track and update the dictionary elements as in \cite{zhuang2019local,wu2018unsupervised}.

\subsection{Algorithm Baselines}
\label{subsec:Algorithm Baselines}
\subsubsection{Deep Clustering based on Spatial continuity (DCS).} As our first baseline, we leverage an inherent spatial continuity of WSIs. Spatially adjacent image patches (tiles) are typically more similar to each other than distant image patches in the slide and therefore should have similar feature representation $\mathcal{Z}$. Hence, we force the model to adopt such behavior by minimizing the distance between feature representations of a specific tile $z_i$ and its overlapping tiles $\mathcal{S}_i$. 
%
\subsubsection{Deep Cluster Assignment (DCA).} The downside of the first baseline is that in some cases, two distant image patches may be visually similar, or there may exist some spatially close patches that are visually different. This introduces noise in the optimization process. To tackle this issue, we can impose cluster membership as in \cite{muhammad2019unsupervised}. 

%
\subsubsection{Deep Embedded Clustering (DEC).} Unlike the second baseline, the objective of our last baseline is not only to determine the clusters but also to learn a meaningful representation of the tiles. Therefore, we consider to jointly learn deep feature representation ($\phi, \psi$) and image clusters $U$. The optimization is performed through the joint minimization of reconstruction loss and the KL divergence to gradually anneal cluster centers by fitting the model to an auxiliary distribution (see \cite{xie2016unsupervised} for details).

\input{files/method_survival}

%% file: files/method_survival.tex
\subsection{Survival Analysis}
\label{subsec:SurvivalAnalysis}

\subsubsection{Clustering and Assignment.} The learned embedding space is assumed to be composed of a limited number of homogeneous clusters. We fit spherical KMeans clustering (SPKM) \cite{skmc05zhon} to the learned latent space with $K$ clusters. As a result, every patch within a patient slide will be assigned to a cluster, $c_k = \argmin_{k \in \{0 \hdots K-1\}} \text{SPKM}(x_i, \mu_k)$.

Our objective is to model the interaction between tumor-related image regions (neighbor patches and clusters). To do so, we define a patient descriptor $h = [h^{C}, h^{T}] \in \mathbb{R}^{N \times (K+K^2)}$ as:

\begin{equation}
     h_{k}^{C} = p(s = k) \quad \text{and} \quad
    h_{j\rightarrow k}^{T} = p(s = k \mid N(s) = j),
    \label{eq:h_features}
\end{equation}

where $s$ is a patch, $h_{k}^{C}$ denotes the probability that a patch belongs to cluster $k$ and $h_{k}^{T}$ is the probability transition between a patch and its neighbors $N(s)$ (e.i. local interactions between clusters within the slide). 

\subsubsection{Survival.} Survival analysis is prone to overfitting as we usually rely on a small patient set and a large number of features. To counter this issue, we first apply forward variable selection \cite{muche2001applied} using log partial likelihood function with tied times \cite{coxtiedtimes77efron}, $\mathcal{L}_{\text{ll}}$, and likelihood-ratio (LR) test to identify the subset of relevant covariates:

\begin{equation}
    \text{LR} = -2[\mathcal{L}_{\text{ll}}( \beta^{\text{new}} \mid h^{\text{new}}) - \mathcal{L}_{\text{ll}}( \beta^{\text{prev}} \mid h^{\text{prev}})].
    \label{eq:forward}
\end{equation}

Here ${(h, \beta)}^{\text{prev}}$ and ${(h, \beta)}^{\text{new}}$ are the previous and new estimated set of covariates, respectively. To validate that the selected covariates do not overfit the patient data, we use leave-one-out cross-validation (LOOCV) on the dataset and predict linear estimators \cite{dai2019cross} as $\hat{\eta}_i = h_i \cdot \beta^{-i}$ and $\hat{\eta} = (\hat{\eta}_1, \hat{\eta}_2, \hdots \hat{\eta}_N)$ to compute C-Index \cite{multivariable96harrel}. Here, $\beta^{-i}$ is estimated on the whole patient set minus patient $i$.

%% file: files/results.tex
\section{Experimental Results}
\label{sec:experimentalResults}

\subsubsection{Dataset.}

We use a set of 660 in-house unlabeled WSIs of CRC stained with hematoxylin and eosin (H\&E). The slides are linked to a total of 374 unique patients diagnosed with adenocarcinoma. The dataset was filtered such that we exclude cases of mucinous adenocarcinoma in which their features are considered independent with respect to standard adenocarcinoma. A set of histopathological features (HFs) is associated with each patient entry (e.i. depth of invasion, pT, etc.). The survival time is defined as the period between resection of the tissue (operation) and the event occurrence (death of the patient). We denote $\mathcal{D}^{S}$ as the dataset that contains slides images and $\mathcal{D}^{S \cap HF}$ as the dataset that contains both information of the HFs and slides for each patient. Note that $|\mathcal{D}^{S \cap HF}| < |\mathcal{D}^{S}|$ as some patients have missing HFs and were excluded.

\subsubsection{Experimental Settings.}
We use ResNet-18 for the encoder where the input layer is updated to support 2 input channels. The latent space has dimensions $d=512$. The decoder is a succession of convolutional layers, ReLUs, and up-samplings (bicubic). The model was trained with the reconstruction loss $\mathcal{L}_{\text{MSE}}$ for 20 epochs with early stopping. We use Adam optimizer $\beta=(0.9, 0.999)$ and learning rate, $lr = 1\mathrm{e}{-3}$. Then, we add $\mathcal{L}_{\text{Divide}}$ for an additional 20 epochs with $\lambda = 1\mathrm{e}{-3}$ and $\tau=0.5$. Finally, we go through 3 additional rounds using $\mathcal{L}_{\text{Rule}}$ while raising the entropy threshold between each round. 

\begin{figure}[t!]
\centering
  \includegraphics[width=\linewidth]{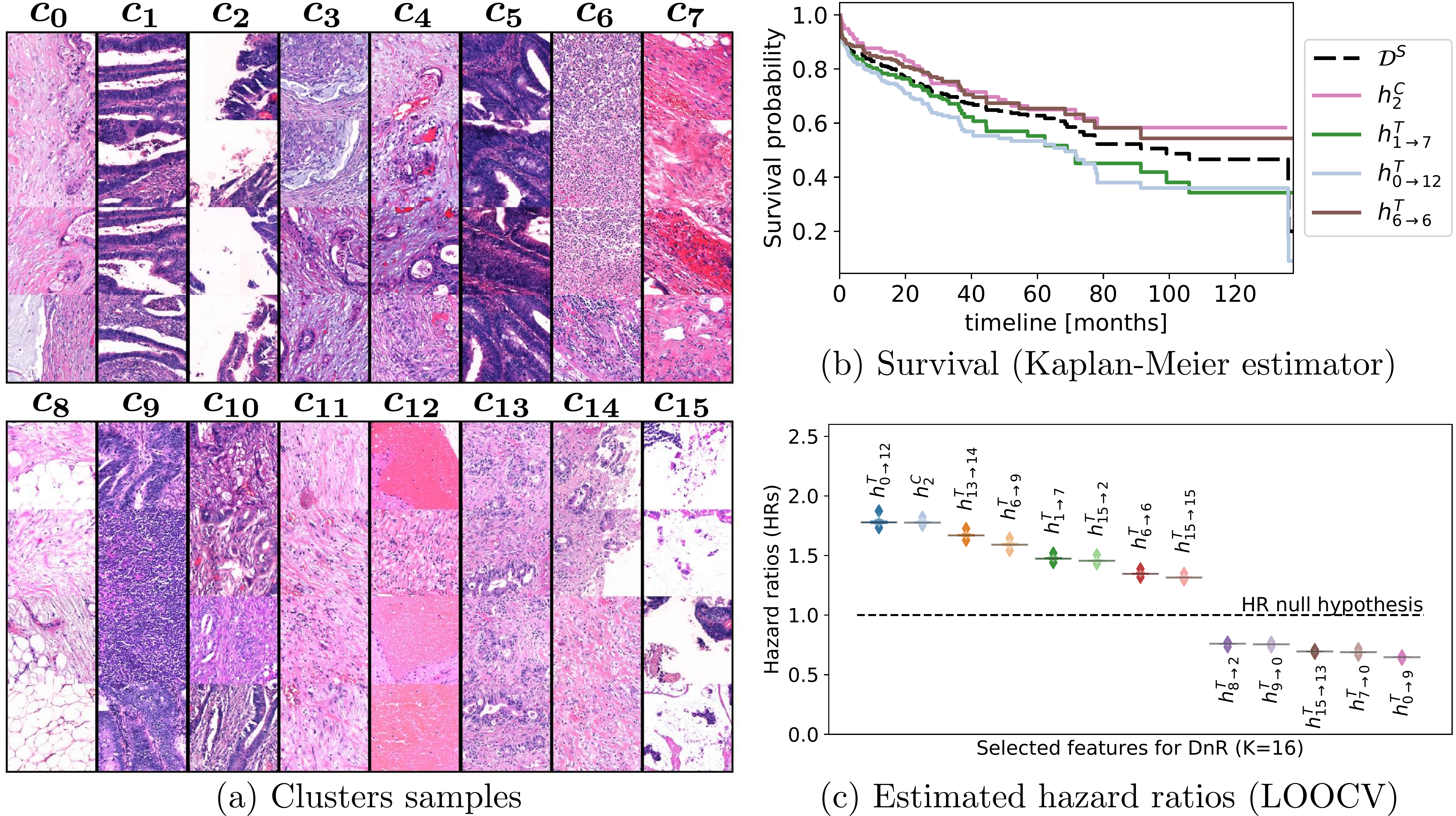}
    \caption{\textbf{Comparison of estimated clusters representation.} (a) Survival results and estimated hazard ratios over LOOCV (b-c). For Kaplan-Meier estimators, we choose a subset of curves that do not overlap too much for better visualization.}
    \label{fig:an_k16_clusters}
\end{figure}

\subsubsection{Clustered Embedding Space.}

We fit SPKM with $K=8$ and $K=16$. The sampled tiles for each cluster are presented in Fig.~\ref{fig:an_k16_clusters}. Clusters demonstrate different tumor and stroma interactions ($c_0$, $c_1$, $c_5$, $c_9$), inflammatory tissues ($c_6$), muscles and large vessels ($c_7$), collagen and small vessels ($c_8$), blood and veins ($c_{11}$) or connective tissues ($c_{12}$). Some clusters do not directly represent the type of tissue but rather the positioning information such as $c_2$, which describe the edge of the WSI.

\input{results/table_mv_score}

\subsubsection{Ablation Study and Survival Analysis Results.}
We build our survival features (Eq.~\ref{eq:h_features}) on top of the predicted clusters, and their contribution is evaluated using Eq.~\ref{eq:forward}. In Tab.~\ref{tab_mv_score}, we observe that our model outperforms previous approaches by a safe $5\%$ margin on C-Index \cite{multivariable96harrel}. The second step of the learning (DnR w/o $\mathcal{L}_{\text{Rule}}$) tends to decrease the prediction score. Such behavior is to be expected as the additional term ($\mathcal{L}_{\text{Divide}}$) will scatter the data and focus on self instance representation. When $\mathcal{L}_{\text{Rule}}$ is then introduced, the model can restructure the embedding by linking similar instances. Also, we observe an augmentation in features, $N_{\text{feat}}$, that achieve statistical relevance for prognosis as we go through our learning procedure (for $K=16$), which proves that our proposed framework can model more subtle patches interactions. We show in Fig.~\ref{fig:an_k16_clusters} the distribution of hazard ratios for all models (from LOOCV) and the Kaplan-Meier estimator \cite{nonparam58kaplanmeier} for a subset of the selected covariates. In the best case, we identify 13 features that contribute to the survival outcome of the patients. For example, the interaction between blood vessels and tumor stroma ($h^{T}_{1\rightarrow7}$) is linked to a lower survival outcome. A similar trend observed in the relation between tumor stroma and connective tissues ($h^{T}_{0\rightarrow12}$).

%% file: results/table_mv_score.tex
\begin{table*}[t!]
\caption{\label{tab_mv_score} Multivariate survival analysis for the proposed approach and baselines. $K$ and $\text{N}_\text{feat}$ denote the number of clusters and the number of features that achieve statistical relevance when performing forward selection ($p < 0.05$). $n$ denotes the number of patient in each set. Brier and Concordance Index are indicators of the performance.}
\centering
\begin{threeparttable}
\begin{tabular}{l l l r d{2.5} r d{2.5}}
\toprule
 &  &  & \multicolumn{2}{c}{$\mathcal{D}^{S \cap HF}\ (n=253)$ } & \multicolumn{2}{c}{$\mathcal{D}^{S}\ (n=374)$} \\
\cmidrule(lr){4-5} \cmidrule(lr){6-7}
Method & $K$ & $\text{N}_\text{feat}$ & Brier \cite{forecast50briern} & \multicolumn{1}{c}{C-Index \cite{multivariable96harrel}} & Brier& \multicolumn{1}{c}{C-Index}\\
\midrule
Histo. features (HFs) &  & 8 & 0.2896 & 0.6076\tnote{***} & - & $-$  \\
\midrule
DCS & 8 & 3 & 0.2840 & 0.5398\tnote{+} & 0.2848 & 0.5562\tnote{**}   \\
DCA${}^\dagger$ \cite{muhammad2019unsupervised} & 8 & 2 & 0.2887 & 0.5452\tnote{**} & 0.2850 & 0.5555\tnote{***}  \\
DEC${}^\dagger$ \cite{xie2016unsupervised} & 8 & 4 & 0.2884 & 0.6089\tnote{**} & 0.2830 & 0.5765\tnote{**} \\
DnR w/o $\mathcal{L}_{\text{Divide}}$, $\mathcal{L}_{\text{Rule}}$ & 8 & 3 & 0.2870 & 0.6070\tnote{*} & 0.2824 & 0.6040\tnote{***}  \\
DnR w/o $\mathcal{L}_{\text{Rule}}$ & 8 & 3 & 0.2828 & 0.5951\tnote{**} & 0.2840 & 0.5919\tnote{***} \\
DnR (ours) & 8 & 4 & 0.2854 & 0.6107\tnote{*} & 0.2832 & 0.6243\tnote{***} \\
\midrule
DCS & 16 & 9 & 0.2934 & 0.6073 & 0.2879 & 0.6464\tnote{***}  \\
DCA${}^\dagger$ \cite{muhammad2019unsupervised} & 16 & 7 & 0.2827 & 0.6246\tnote{+} & 0.2852 & 0.6322\tnote{**}  \\
DEC${}^\dagger$ \cite{xie2016unsupervised} & 16 & 7 & \textbf{0}.\textbf{2758} & 0.6410\tnote{**} & 0.2763 & 0.6426\tnote{***}  \\
DnR w/o $\mathcal{L}_{\text{Divide}}$, $\mathcal{L}_{\text{Rule}}$ & 16 & 5 & 0.2819 & 0.6364\tnote{*} & 0.2795 & 0.6324\tnote{***}  \\
DnR w/o $\mathcal{L}_{\text{Rule}}$ & 16 & 10 & 0.3006 & 0.6207\tnote{+} & 0.2934 & 0.6468\tnote{***} \\
DnR (ours) & 16 & 13 & 0.2849 & \textbf{0}.\textbf{6736}\tnote{**} & \textbf{0}.\textbf{2725} & \textbf{0}.\textbf{6943}\tnote{***} \\
\bottomrule
\end{tabular}
\begin{tablenotes}

\small
\item[$\dagger$]\ Autoencoder is replaced with the self-supervised objective function.
\item[+]\ $p<0.1$; $^{*}\ p<0.05$; $^{**}\ p<0.01$; $^{***}\ p<0.001$ (log-rank test).
\end{tablenotes}
\end{threeparttable}
\end{table*}

%% file: files/conclusion.tex
\section{Conclusion}
\label{sec:conclusion}
We have proposed a self-supervised learning method that offers a new approach to learn histopathological patterns within cancerous tissue regions. Our model presents a novel way to model the interactions between tumor-related image regions and tackles the inherent overfitting problem to predict patient outcome. Our method surpasses all previous baseline methods and histopathological features and achieves state-of-the-art results, i.e., in C-Index without any data-specific annotation. Ablation studies also show the importance of different components of our method and the relevance of combining them. We envision the broad application of our approach for clinical prognostic stratification improvement.


%% file: files/appendix.tex
\clearpage
\section{Additional Figures and Results}

\begin{figure}[h!]
     \centering
     \begin{subfigure}[b]{0.3\textwidth}
         \centering
         \includegraphics[width=\textwidth]{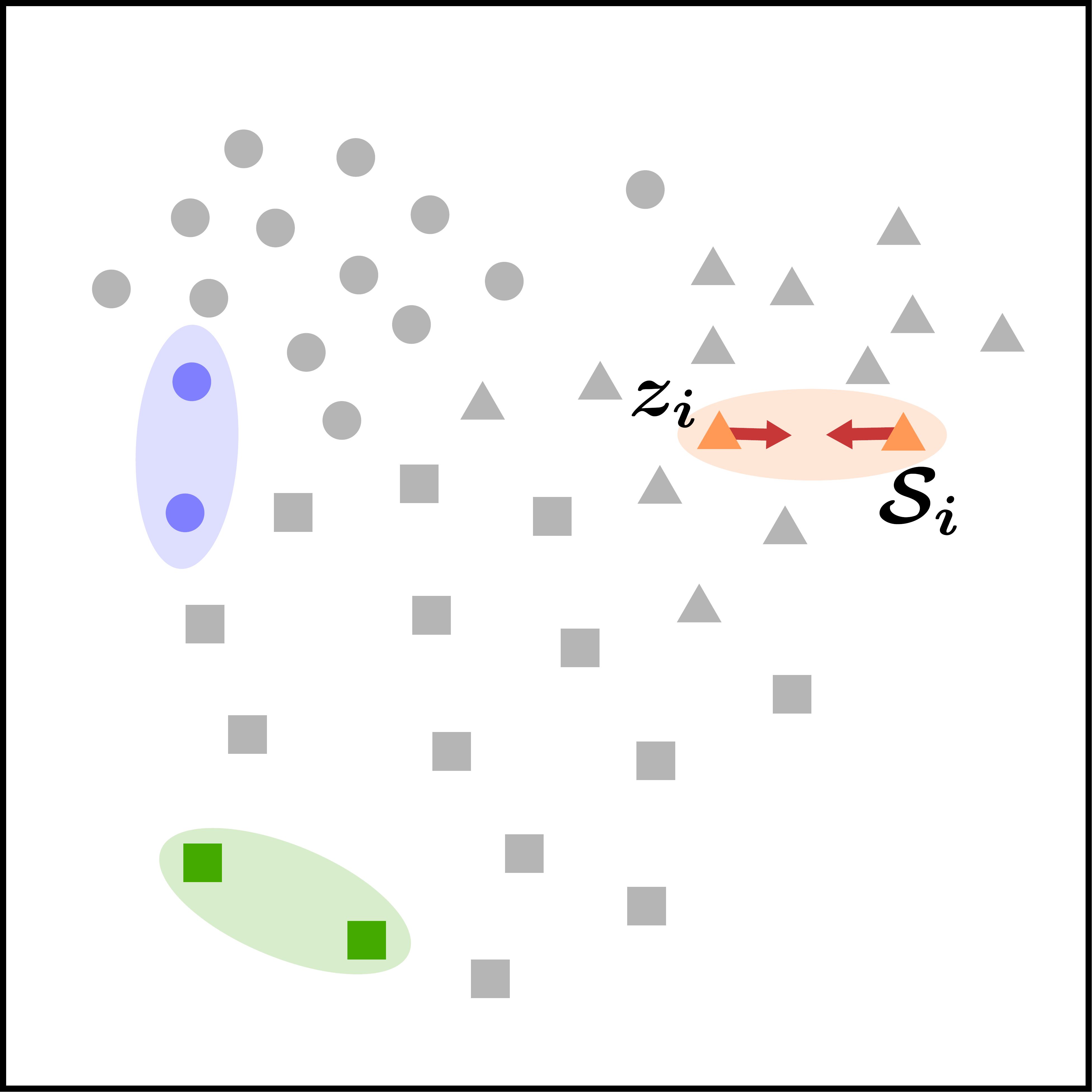}
         \label{fig:clustering_l2}
     \end{subfigure}
     \begin{subfigure}[b]{0.3\textwidth}
         \centering
         \includegraphics[width=\textwidth]{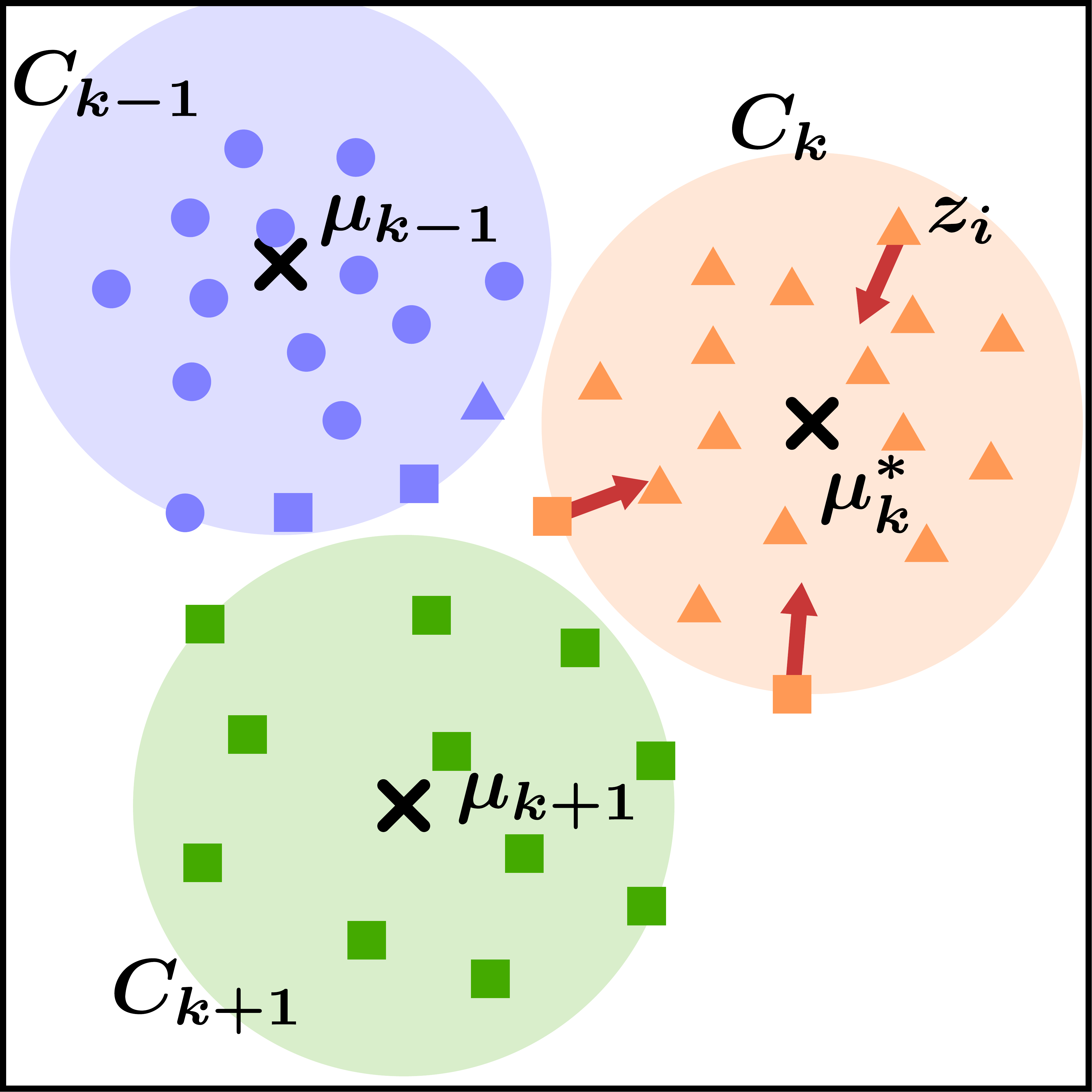}
         \label{fig:clustering_knn}
     \end{subfigure}
     \begin{subfigure}[b]{0.3\textwidth}
         \centering
         \includegraphics[width=\textwidth]{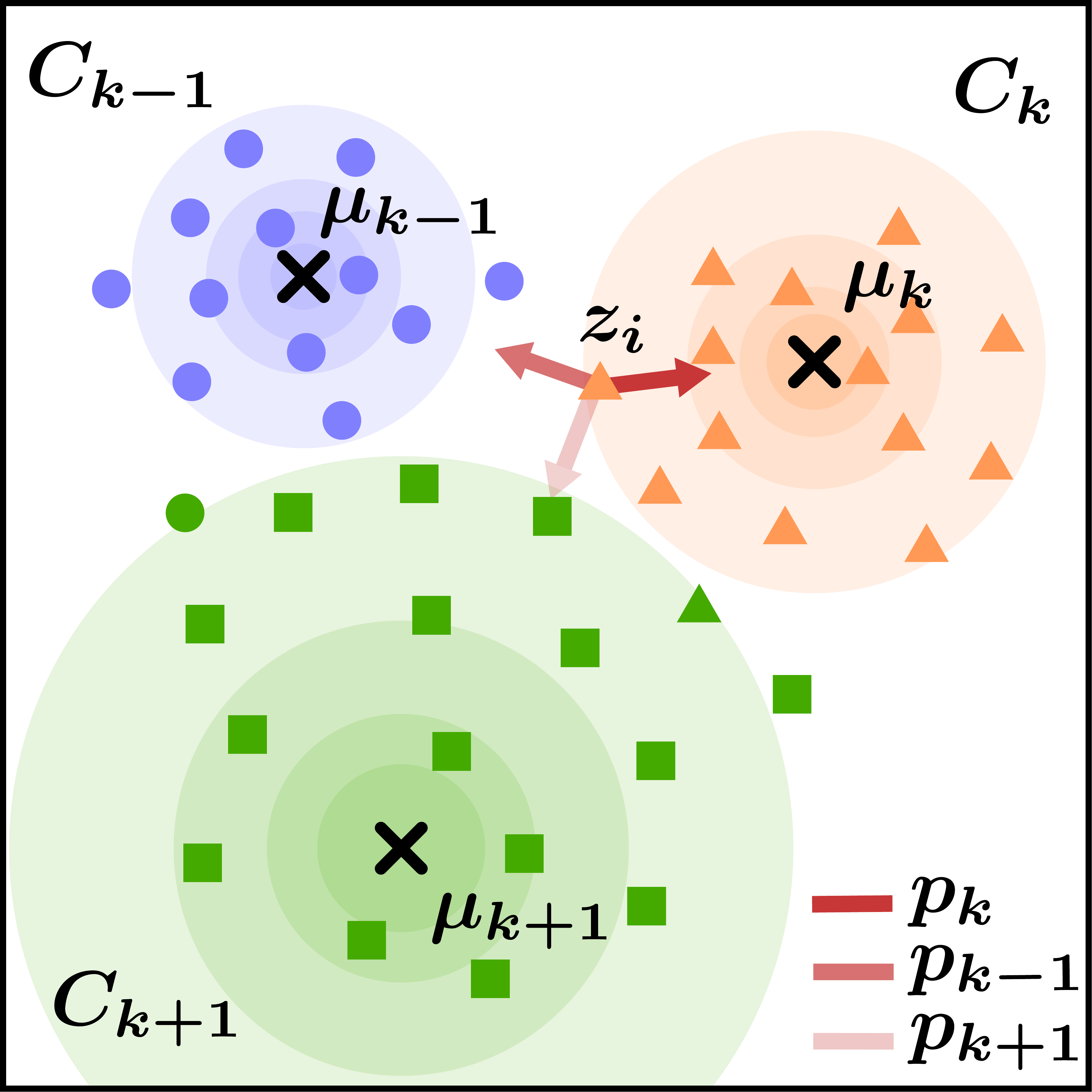}
         \label{fig:clustering_dec}
     \end{subfigure}
        \caption{
        Schematic of proposed lower-dimensional embedding representations. From left to right - DCS, DCA, DEC.
        }
        \label{fig:clustering}
\end{figure}

\input{results/table_he_rgb.tex}
\input{results/table_hr_bern.tex}

\clearpage

%% file: results/table_he_rgb.tex
\begin{table*}
\caption{\label{tab_mv_score_sup} Multivariate survival analysis comparison between self-supervised when training $\text{RGB} \rightarrow \text{RGB}$ and $\text{HE} \rightarrow \text{RGB}$. We can observe that the model performs better when we impose the color conversion from the HE space.}
\centering
\begin{threeparttable}
\begin{tabular}{l l l r d{2.5} r d{2.5}}
\toprule
 &  &  & \multicolumn{2}{c}{$\mathcal{D}^{S \cap HF}\ (n=253)$ } & \multicolumn{2}{c}{$\mathcal{D}^{S}\ (n=374)$} \\
\cmidrule(lr){4-5} \cmidrule(lr){6-7}
Method & $K$ & $\text{N}_\text{feat}$ & Brier & \multicolumn{1}{c}{C-Index} & Brier& \multicolumn{1}{c}{C-Index}\\
\midrule
$\text{MSE}_{\text{RGB} \rightarrow \text{RGB}}$ & 8 & 2 & 0.2848 & 0.5272\tnote{**} & 0.2859 & 0.5110\tnote{*}  \\
$\text{MSE}_{\text{HE} \rightarrow \text{RGB}}$ & 8 & 3 & 0.2870 & 0.6070\tnote{*} & 0.2824 & 0.6040\tnote{***}  \\
\midrule
$\text{MSE}_{\text{RGB} \rightarrow \text{RGB}}$ & 16 & 0 & 0.2893 & 0.5000 & 0.2896 & 0.5000   \\
$\text{MSE}_{\text{HE} \rightarrow \text{RGB}}$ & 16 & 5 & \textbf{0}.\textbf{2819} & \textbf{0}.\textbf{6364}\tnote{*} & \textbf{0}.\textbf{2795} & \textbf{0}.\textbf{6324}\tnote{***}   \\
\bottomrule
\end{tabular}
\begin{tablenotes}

\small
\item[+]\ $p<0.1$; $^{*}\ p<0.05$; $^{**}\ p<0.01$; $^{***}\ p<0.001$ (log-rank test).
\end{tablenotes}
\end{threeparttable}
\end{table*}

%% file: results/table_hr_bern.tex
\begin{table*}
\caption{\label{tab_hr_bern} Hazard ratios (HRs) with confidence intervals (CIs) based on histopathological features for $n=374$ patients with adenocarcinoma (241 right censored samples). Results are given for the univariate Cox model. If entry is non-binary, we apply one-vs.-all test.}
\centering
\begin{threeparttable}
\begin{tabular}{llrr}
\toprule
Characteristics & Subcategories & HR (95\% CI) & $p$-value \\
\midrule
Gender ($n=374$) & Male ($n=220$) &  &  \\
   & Female ($n=154$) & 0.90 (0.64 - 1.28) & 0.5139 \\
\midrule
T category ($n=373$) & pT1-2 ($n=78$) &  &  \\
   & pT3-4 ($n=295$) & 1.67 (1.00 - 2.79) & 0.0479\tnote{*} \\
\midrule
N category ($n=366$) & pN0 ($n=186$) &  &  \\
   & pN1-2 ($n=180$)  & 2.65 (1.83 - 3.82) & $<$ 0.0001\tnote{*}\\
\midrule
M category ($n=374$) & pM0 ($n=326$) &  &  \\
   & pM1 ($n=48$) & 1.76 (1.11 - 2.78) & 0.0162\tnote{*} \\
\midrule
Tumor grade ($n=369$) & G1-2 ($n=324$) &  &  \\
   & G3 ($n=45$) & 1.54 (0.96 - 2.46) & 0.0716\\
\midrule
Lymphatic invasion ($n=351$) & L0 ($n=141$) &  &  \\
   & L1 ($n=210$) & 3.35 (2.14 - 5.23) & $<$ 0.0001\tnote{*}\\
\midrule
Vascular invasion ($n=352$) & V0 ($n=193$) &  &  \\
   & V1-2 ($n=159$) & 1.52 (1.07 - 2.16) & 0.0198\tnote{*} \\
\midrule
Tumor pushing ($n=276$) & $<25\%$ ($n=109$) &  &  \\
   & $\geq25\%$ ($n=167$) & 0.59 (0.40 - 0.87) & 0.0077\tnote{*}\\
\midrule
Tumor location ($n=352$) & Left ($n=164$) &  1.28 (0.90 - 1.83) & 0.1702 \\
   & Rectum ($n=58$) & 0.83 (0.50 - 1.36) & 0.4535 \\
   & Right ($n=130$) & 0.85 (0.58 - 1.24) & 0.3967 \\
\midrule
TNM stage ($n=372$) & I ($n=64$) & 0.45 (0.24 - 0.83) & 0.0109\tnote{*} \\
   & II ($n=114$) & 0.50 (0.33 - 0.75) & 0.0010\tnote{*} \\
   & III ($n=118$) & 1.57 (1.11 - 2.22) & 0.0102\tnote{*} \\
   & IV ($n=76$) & 2.03 (1.39 - 2.96)  & 0.0003\tnote{*} \\
\bottomrule
\end{tabular}
\begin{tablenotes}
  \item[*] Indicates statistical relevance ($\alpha = 0.05$).
\end{tablenotes}
\end{threeparttable}
\end{table*}

%% file: paper616.bbl
\begin{thebibliography}{10}
\providecommand{\url}[1]{\texttt{#1}}
\providecommand{\urlprefix}{URL }
\providecommand{\doi}[1]{https://doi.org/#1}

\bibitem{forecast50briern}
Brier, G.W.: Verification of forecasts expressed in terms of probability.
  Monthly weather review  \textbf{78}(1), ~1--3 (1950)

\bibitem{chen2020simple}
Chen, T., Kornblith, S., Norouzi, M., Hinton, G.: A simple framework for
  contrastive learning of visual representations. arXiv preprint
  arXiv:2002.05709  (2020)

\bibitem{dai2019cross}
Dai, B., Breheny, P.: Cross validation approaches for penalized cox regression.
  arXiv preprint arXiv:1905.10432  (2019)

\bibitem{dercksen2019dealing}
Dercksen, K., Bulten, W., Litjens, G.: Dealing with label scarcity in
  computational pathology: A use case in prostate cancer classification. arXiv
  preprint arXiv:1905.06820  (2019)

\bibitem{coxtiedtimes77efron}
Efron, B.: The efficiency of cox's likelihood function for censored data.
  Journal of the American statistical Association  \textbf{72}(359),  557--565
  (1977)

\bibitem{fouad2017unsupervised}
Fouad, S., Randell, D., Galton, A., Mehanna, H., Landini, G.: Unsupervised
  morphological segmentation of tissue compartments in histopathological
  images. PloS one  \textbf{12}(11),  e0188717 (2017)

\bibitem{multivariable96harrel}
Harrell~Jr, F.E., Lee, K.L., Mark, D.B.: Multivariable prognostic models:
  issues in developing models, evaluating assumptions and adequacy, and
  measuring and reducing errors. Statistics in medicine  \textbf{15}(4),
  361--387 (1996)

\bibitem{he2020momentum}
He, K., Fan, H., Wu, Y., Xie, S., Girshick, R.: Momentum contrast for
  unsupervised visual representation learning. In: Proceedings of the IEEE/CVF
  Conference on Computer Vision and Pattern Recognition. pp. 9729--9738 (2020)

\bibitem{he16}
He, K., Zhang, X., Ren, S., Sun, J.: Deep residual learning for image
  recognition. In: Proceedings of the IEEE conference on computer vision and
  pattern recognition. pp. 770--778 (2016)

\bibitem{muche2001applied}
Hosmer~Jr, D.W., Lemeshow, S., May, S.: Applied survival analysis: regression
  modeling of time-to-event data, vol.~618. John Wiley \& Sons (2011)

\bibitem{nonparam58kaplanmeier}
Kaplan, E.L., Meier, P.: Nonparametric estimation from incomplete observations.
  Journal of the American statistical association  \textbf{53}(282),  457--481
  (1958)

\bibitem{kather19dataset}
Kather, J.N., Halama, N., Marx, A.: {100,000 histological images of human
  colorectal cancer and healthy tissue} (Apr 2018).
  \doi{10.5281/zenodo.1214456}, \url{https://doi.org/10.5281/zenodo.1214456}

\bibitem{kather2019predicting}
Kather, J.N., Krisam, J., Charoentong, P., Luedde, T., Herpel, E., Weis, C.A.,
  Gaiser, T., Marx, A., Valous, N.A., Ferber, D., et~al.: Predicting survival
  from colorectal cancer histology slides using deep learning: A retrospective
  multicenter study. PLoS medicine  \textbf{16}(1),  e1002730 (2019)

\bibitem{li2018graph}
Li, R., Yao, J., Zhu, X., Li, Y., Huang, J.: Graph cnn for survival analysis on
  whole slide pathological images. In: International Conference on Medical
  Image Computing and Computer-Assisted Intervention. pp. 174--182. Springer
  (2018)

\bibitem{macenko09stain}
{Macenko}, M., {Niethammer}, M., {Marron}, J.S., {Borland}, D., {Woosley},
  J.T., {Xiaojun Guan}, {Schmitt}, C., {Thomas}, N.E.: A method for normalizing
  histology slides for quantitative analysis. In: 2009 IEEE International
  Symposium on Biomedical Imaging: From Nano to Macro. pp. 1107--1110 (June
  2009)

\bibitem{moriya2018unsupervised}
Moriya, T., Roth, H.R., Nakamura, S., Oda, H., Nagara, K., Oda, M., Mori, K.:
  Unsupervised pathology image segmentation using representation learning with
  spherical k-means. In: Medical Imaging 2018: Digital Pathology. vol. 10581,
  p. 1058111. International Society for Optics and Photonics (2018)

\bibitem{muhammad2019unsupervised}
Muhammad, H., Sigel, C.S., Campanella, G., Boerner, T., Pak, L.M., B{\"u}ttner,
  S., IJzermans, J.N., Koerkamp, B.G., Doukas, M., Jarnagin, W.R., et~al.:
  Unsupervised subtyping of cholangiocarcinoma using a deep clustering
  convolutional autoencoder. In: International Conference on Medical Image
  Computing and Computer-Assisted Intervention. pp. 604--612. Springer (2019)

\bibitem{vahadane16stain}
{Vahadane}, A., {Peng}, T., {Sethi}, A., {Albarqouni}, S., {Wang}, L., {Baust},
  M., {Steiger}, K., {Schlitter}, A.M., {Esposito}, I., {Navab}, N.:
  Structure-preserving color normalization and sparse stain separation for
  histological images. IEEE Transactions on Medical Imaging  \textbf{35}(8),
  1962--1971 (Aug 2016)

\bibitem{wu2018unsupervised}
Wu, Z., Xiong, Y., Yu, S.X., Lin, D.: Unsupervised feature learning via
  non-parametric instance discrimination. In: Proceedings of the IEEE
  Conference on Computer Vision and Pattern Recognition. pp. 3733--3742 (2018)

\bibitem{xie2016unsupervised}
Xie, J., Girshick, R., Farhadi, A.: Unsupervised deep embedding for clustering
  analysis. In: International conference on machine learning. pp. 478--487
  (2016)

\bibitem{skmc05zhon}
Zhong, S.: Efficient online spherical k-means clustering. In: Proceedings. 2005
  IEEE International Joint Conference on Neural Networks, 2005. vol.~5, pp.
  3180--3185. IEEE (2005)

\bibitem{zhu2017wsisa}
Zhu, X., Yao, J., Zhu, F., Huang, J.: Wsisa: Making survival prediction from
  whole slide histopathological images. In: Proceedings of the IEEE Conference
  on Computer Vision and Pattern Recognition. pp. 7234--7242 (2017)

\bibitem{zhuang2019local}
Zhuang, C., Zhai, A.L., Yamins, D.: Local aggregation for unsupervised learning
  of visual embeddings. In: Proceedings of the IEEE International Conference on
  Computer Vision. pp. 6002--6012 (2019)

\end{thebibliography}
